\documentclass[]{spie}  
\usepackage[]{graphicx}
\usepackage{caption}
\usepackage{subcaption}
\usepackage{doi}
\usepackage[none]{hyphenat}
\usepackage{amsmath}
\usepackage{import}
\usepackage{amsmath,latexsym,amsfonts}
\usepackage{siunitx}
\usepackage{color}
\graphicspath{{figs/}}

\title{GRAVITY acquisition camera: characterization results} 
\author{Narsireddy Anugu\supit{*a}, Paulo Garcia\supit{a}, Antonio Amorim\supit{b}, Erich Wiezorrek\supit{c}, Ekkehard Wieprecht\supit{c},   Frank Eisenhauer\supit{c},  Thomas Ott\supit{c}, Oliver Pfuhl\supit{c}, Paulo Gordo\supit{b}, Guy Perrin\supit{d}, Wolfgang Brandner\supit{e}, 
Christian Straubmeier\supit{f} and Karine Perraut\supit{g}
\skiplinehalf
\supit{a}Universidade do Porto, Faculdade de Engenharia,  Unidade FCT CENTRA, Rua Dr. Roberto Frias, s/n, P-4200-465 Porto, Portugal; 
\supit{b}CENTRA-SIM and FCUL - Edificio C8, Campo Grande, P1749016 Lisboa, Portugal; \\
\supit{c}Max Planck Institute for extraterrestrial Physics, PO Box 1312, Giessenbachstr., 85741 Garching, Germany; \\
 \supit{d}Observatoire de Paris Meudon, Paris, France;\\ \supit{e}MaxPlanck-Institut fur Astronomie, Heidelberg, Germany;  \\ \supit{f}Universit\"{a}t zu K\"{o}ln Germany;\\  \supit{g}Institut de Plan\'{e}tologie et d\'{}Astrophysique de Grenoble (IPAG), Grenoble, France
}

\authorinfo{*narsireddy.anugu@fe.up.pt}
\begin{document} 
\maketitle

\begin{abstract}
GRAVITY acquisition camera implements four optical functions to track multiple beams of Very Large Telescope Interferometer (VLTI): a) pupil tracker: a $2 \times 2$ lenslet images four pupil reference lasers mounted on the spiders of telescope secondary mirror; b) field tracker: images science object; c) pupil imager: reimages telescope pupil; d) aberration tracker: images a Shack-Hartmann. The estimation of beam stabilization parameters from the acquisition camera detector image is carried out, for every 0.7 s, with a dedicated data reduction software. The measured parameters are used in: a) alignment of GRAVITY with the VLTI; b) active pupil and field stabilization; c)  defocus correction and engineering purposes. The instrument is now successfully operational on-sky in closed loop. The relevant data reduction and on-sky characterization results are reported.
\end{abstract}



\section{INTRODUCTION}

GRAVITY is a high precision narrow angle astrometry and  interferometric imaging instrument~\cite{Eisenhauer2011}. It has been built for the Very Large Telescope Interferometer (VLTI) of the European Southern Observatory and was born with the goal of monitoring the stellar sources in the vicinity of the Galactic Center super massive black hole and the actual emission of the black hole close environment environment. It combines four beams (in the K-band) coherently of either Unit Telescopes (UTs) or Auxiliary Telescopes (ATs), delivering an astrometry of $\sim 10$ micro-arcseconds ($\mu$as) and a angular resolution of around $\sim 4$ milli-arcseconds (mas).

The main error sources for the GRAVITY astrometric  measurements are the atmospheric turbulence and the error in the baseline length ($B$) between any two telescopes. The GRAVITY-Coud\'e Infrared Adaptive Optics (CIAO) measures and corrects wavefront aberrations induced by the atmospheric turbulence in the light path from  the sky to the Coud\'e laboratory~\cite{Kendrew2012}. However the CIAO corrections do not include the VLTI tunnel seeing induced fast moving tip-tilts. These are tracked by launching external lasers at the telescopes (at the star separator) and imaging them in the GRAVITY beam combiner~\cite{Pfuhl2014}. The measured tip-tilts are corrected by dedicated tip-tilt-piston mirrors. But the above corrections involve  tip-tilt residuals of more than 10\,mas. These tip-tilt errors limit the performance of GRAVITY in two ways~\cite{Lacour2014a}: a) efficiency of star light  injection into the single mode fibers which are used to transport the light to the integrated optics~\cite{Jocou2010} to make the beam combination; b) unwanted tip-tilt error causes additional astrometric error. 

The factor limiting the precision  in measuring the accurate baseline (RMS error of a few mm for a $\sim 100$\,m baseline) length between two telescopes is  pupil position errors. These errors are  mainly due to the misalignments and  temperature gradients. In between the telescope pupil and the fiber-fed beam combiner, there exists several optical pupils that are under motion due to the optical train vibrations while tracking the object of interest. The typical lateral and longitudinal pupil position errors experienced by the VLTI are around 5\% in the pupil diameter and  1\,m, respectively,  for the 80\,mm beam. Eq. \ref{Chap2:EqAstrometricError}  gives the astrometric error~\cite{Lacour2015c} associated with the given tip-tilt error ($\Delta \alpha$) and pupil lateral position ($\Delta L$) errors. For example, for 10\,mas tip-tilt error and 0.4\,m (5\% of 8~m pupil) lateral pupil position error, the astrometric error becomes 40\,$\mu$as for a single beam combiner 

\begin{equation}\label{Chap2:EqAstrometricError}
\sigma = \dfrac{\Delta \alpha \times \Delta L}{ B} .
\end{equation}

The CIAO measures and corrects the incoming distorted wavefronts with respect to a flat wavefront generated in the Coud\'e laboratory. However, there are many optics and tunnel seeing  in between the CIAO and the GRAVITY beam combiner. They will  also introduce additional wavefront aberrations to the incoming beam.  The  quasi-static wavefront aberrations and the non-common path errors that exist between the CIAO and the GRAVITY beam combiner also contribute to the astrometric error~\cite{Lacour2014a}.

Therefore a beam stabilization system is required to address the above issues. The GRAVITY acquisition camera has been built to image and analyze the VLTI beams and  to: a) enable the stabilization of star light injection into the single mode fibers; b) minimize the astrometric errors induced by the field and the pupil errors.  

\section{Acquisition camera}

\begin{figure}[h!] 
	\centering{
		\def\svgwidth{\textwidth}
		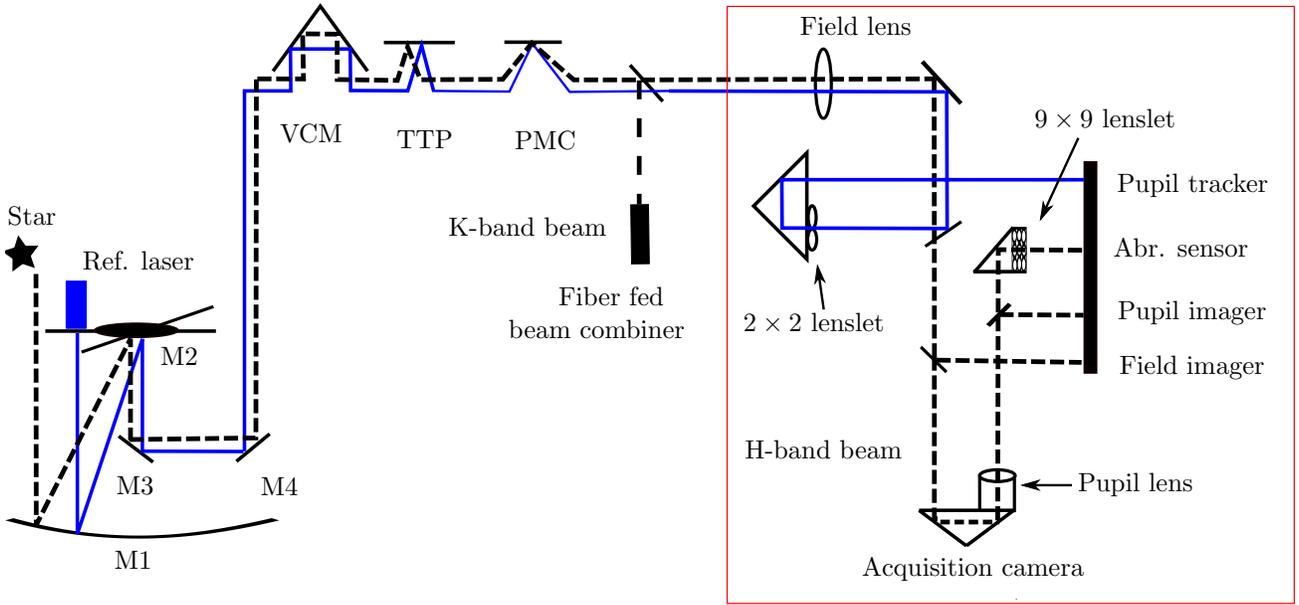 
		\vspace{11pt}
		\caption{\label{Chap3:Fig1} The acquisition camera working principle and optical layout (for clarity only one telescope beam case is shown). The astrophysical target's K-band beam is used for science measurements and the H-band beam is used for  beam stabilization. The H-band beam feeds the  three imaging modes of the acquisition camera: aberration sensor, pupil imager, field imager. The pupil tracker images external pupil reference laser light (\SI{1.2}{\micro m}).  The box in red color encloses the acquisition camera optics.}
	}
\end{figure}

Figure~\ref{Chap3:Fig1} presents the optical layout of the acquisition camera for one telescope beam. The acquisition camera provides four optical imaging modes: a) field imager -- images and tracks the field (tip-tilt) of input beams; b) pupil tracker -- tracks the telescope pupil motions by imaging  four pupil beacons that are installed on the spiders of secondary mirror (M2) of the telescope; c) aberration sensor -- implements four $9\times9$ lenslets to sense the incoming distorted wavefronts; d) pupil imager -- images telescope pupil, which is used for monitoring the pupil visually during the observations.  Pupil tracker images four pupil guiding laser beacons of \SI{1.2}{\micro m}  wavelength with a $2\times2$ lenslet. The other three imaging modes image astrophysical targets in the H-band. The acquisition  camera uses a $2048 \times 2048$ pixel Hawaii-2RG detector to image all four input telescope beams. The optical design of the camera~\cite{Amorim2012} and its optical alignment and integration~\cite{Gordo2014} are presented in the previous SPIE proceedings.

To extract the beam stabilization parameters from the detector images, a dedicated software has been developed based on the image analysis methods presented in Anugu et al. (2014)~\cite{Anugu2014S}. The software was written in C and C++ using CLIP library~\cite{BallesterP.2008} and integrated in the GRAVITY observational software~\cite{burtscher2014gravity}. This software works on the instrument workstation and does the image analysis in two steps. First, it reads the acquisition camera detector image from the instrument shared memory for every 0.7\,s (detector frame rate). Second, it evaluates the beam's tip-tilts, pupil shifts and the wavefront aberrations by analyzing the input detector image. The evaluated parameters are written to the instrument database. They feed the beam stabilization which uses: a) the tip-tilt and piston (TTP); b)the pupil motion controller (PMC)~\cite{Pfuhl2014} and; c) the Variable Curvature Mirror (VCM)~\cite{Ferrari2003}. By stabilizing the field, it enables the injection of the K-band light of the astrophysical targets into single mode fibers which transport the beams towards the interferometric beam combiner.  The image acquisition, analysis and beam correction are carried out on-line in the instrument workstation in parallel to the observations for all four telescopes.    

\section{Characterization results}

\begin{figure}[h!]
	\parbox[t]{11pt}{\rotatebox{90}{meter}} 
	\begin{tabular}
		{@{}c@{}} 
		\begin{subfigure}{0.47\textwidth}
			\centering
			\includegraphics[width=\textwidth]{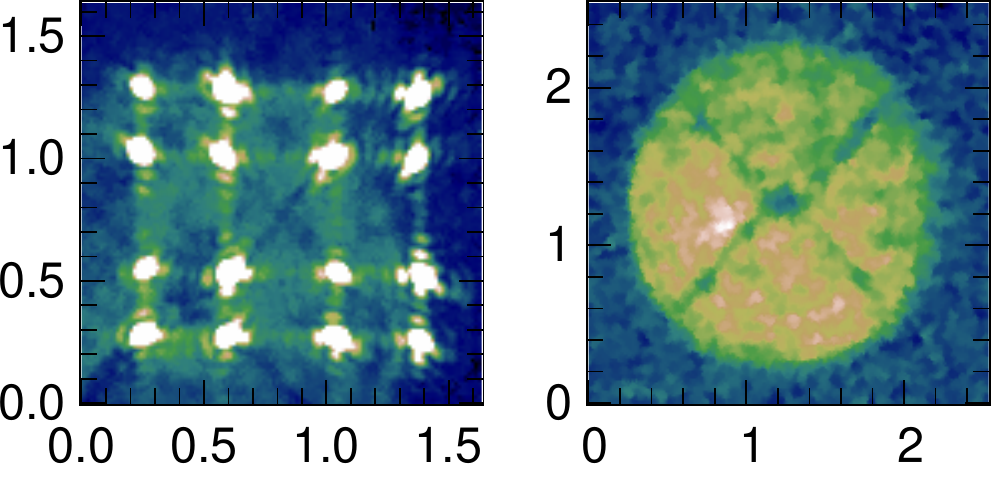}      meter       
		\end{subfigure}
	\end{tabular} \hspace{5pt} \rotatebox{90}{arc-sec}
	\begin{tabular}{
			@{}c@{}}
		\begin{subfigure}{0.47\textwidth}
			\centering
			\includegraphics[width=\textwidth]{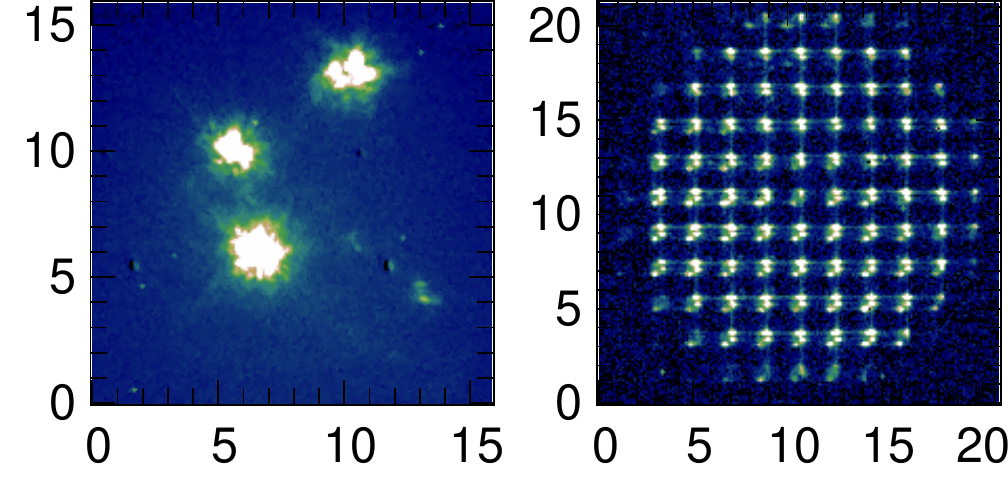} arc-sec
		\end{subfigure}
	\end{tabular} 
	\vspace{11pt}
	\caption{\label{fig9} On-sky measurements of the acquisition camera imaging modes (left to right):  pupil tracker,  pupil imager,   field tracker and  aberration sensor. The astrophysical target is $\theta^1$ Ori A (North up and East on right). The aberration sensor image is rotated counter clockwise to the field image due to the mirror reflection.} 
\end{figure}

GRAVITY passed Preliminary Verification Europe tests (PAE) in  early 2015 and was moved in August 2015 to the Paranal Observatory where it got its first light in November 2015. Figure~\ref{fig9} presents the acquisition camera detector image acquired on-sky with the ATs.

\subsection{\label{LAB}Characterization for PAE}

During the PAE, the acquisition camera was characterized using the GRAVITY calibration unit~\cite{Blind2014}.  The calibration unit is a GRAVITY subsystem which provides  laboratory beams and artificial  stars  to test  GRAVITY functionalities. It generates two artificial stars and four pupil guiding reference beams to test the acquisition camera imaging modes.   

\subsubsection{Field tracking}\label{FI}

Characterization of the field tracker is accomplished in three fronts: a) RMS and absolute tracking accuracy; b) RMS accuracy as a function of target magnitude; c) target flux injection into the fibers.  To characterize the absolute accuracy of the field tracker, known tip-tilts ($\theta_{\rm i}$) are applied to the incoming beams by manipulating the TTP device and measured backm ($\theta_{\rm o}$), using the field tracker function.  

\begin{figure}[h!]
	\parbox[t]{18pt}{\rotatebox{90}{ \hspace{-0.8cm}$|\theta_{\rm i}$ - $\theta_{\rm o}|$ (mas)  }} 
	\begin{tabular}
		{@{}c@{}} 
		\begin{subfigure}{0.3\textwidth}
			\centering
			\includegraphics[width=\textwidth]{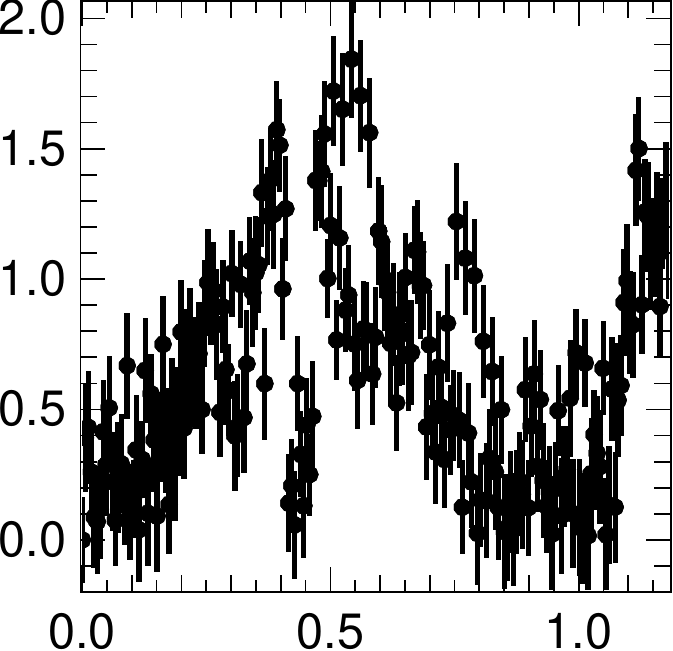}    $\theta_{\rm i}$  (arcsec)
		\end{subfigure}
	\end{tabular} \hspace{5pt} \rotatebox{90}{\hspace{-0.5cm}RMS error (mas)}
	\begin{tabular}{
			@{}c@{}}
		\begin{subfigure}{0.3\textwidth}
			\centering
			\includegraphics[width=\textwidth]{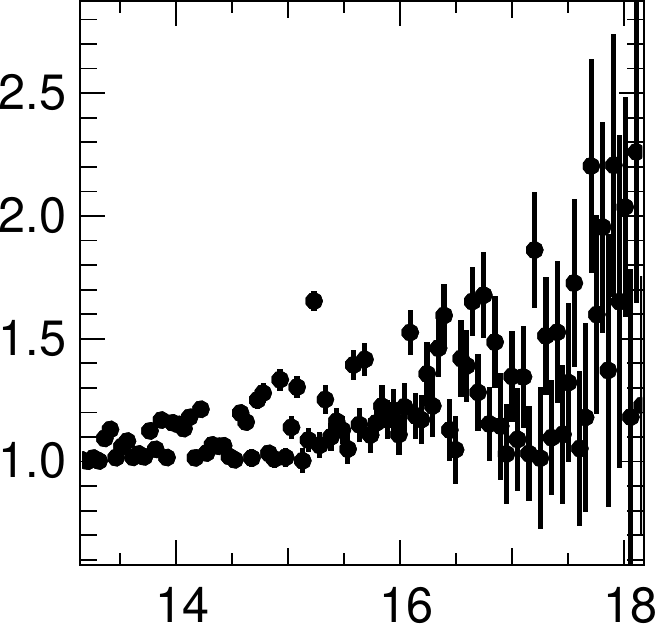} Target magnitude
		\end{subfigure}
			\parbox[t]{11pt}{\rotatebox{90}{\hspace{-1.5cm} Normalized flux in the fiber}}
		\begin{subfigure}{0.27\textwidth}
			\centering
			\includegraphics[width=\textwidth]{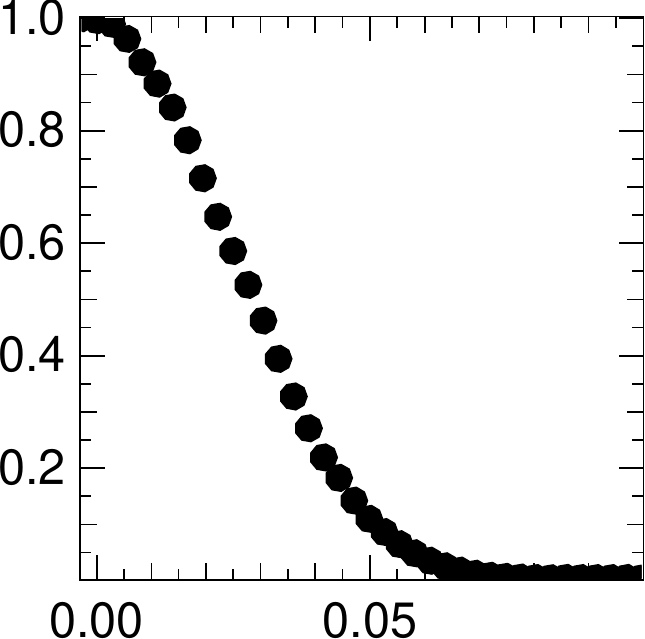} Star shift from the fiber center (arcsec)
		\end{subfigure} 
	\end{tabular} 
	\vspace{11pt}
	\caption{\label{Chap7:fig5} Performance of the field tracker in the laboratory. Left: the absolute field tracking error $|\theta_{\rm i}-\theta_{\rm o}|$  as a function of true object position 	$\theta_{\rm i}$. Middle: the RMS error of the field tracking as a function of target magnitude.  Right: Normalized star's flux injection into the fiber as function of tip-tilt/star shift from the fiber center. } 
\end{figure}

Figure~\ref{Chap7:fig5} (left panel) presents the absolute tracking accuracy and is around $\sim 2$\,mas and the RMS error is $\sim 1$\,mas. These measurements are carried out with a "star" with an H-band magnitude of 15.  Figure~\ref{Chap7:fig5} (middle panel) presents the RMS error of the field tracker as a function of "star" magnitude. These "stars" are realized by the Calibration Unit and by varying the voltage of the lamp.  Figure~\ref{Chap7:fig5} (right panel)  presents how the flux injection in the fiber is reduced. By enabling the field stabilization, the coupling efficiency of the fiber is maintained around 80\%. 

\subsubsection{Lateral pupil tracking}

\begin{figure}[h!]
	\centering
	\parbox[t]{11pt}{\rotatebox{90}{\hspace{-0.8cm} $|L_{\rm {x_0}} - L_{\rm x}|$ (mm) }} 
	\begin{tabular}
		{@{}c@{}} 
		\begin{subfigure}{0.3\textwidth}
			\centering
			\includegraphics[width=\textwidth]{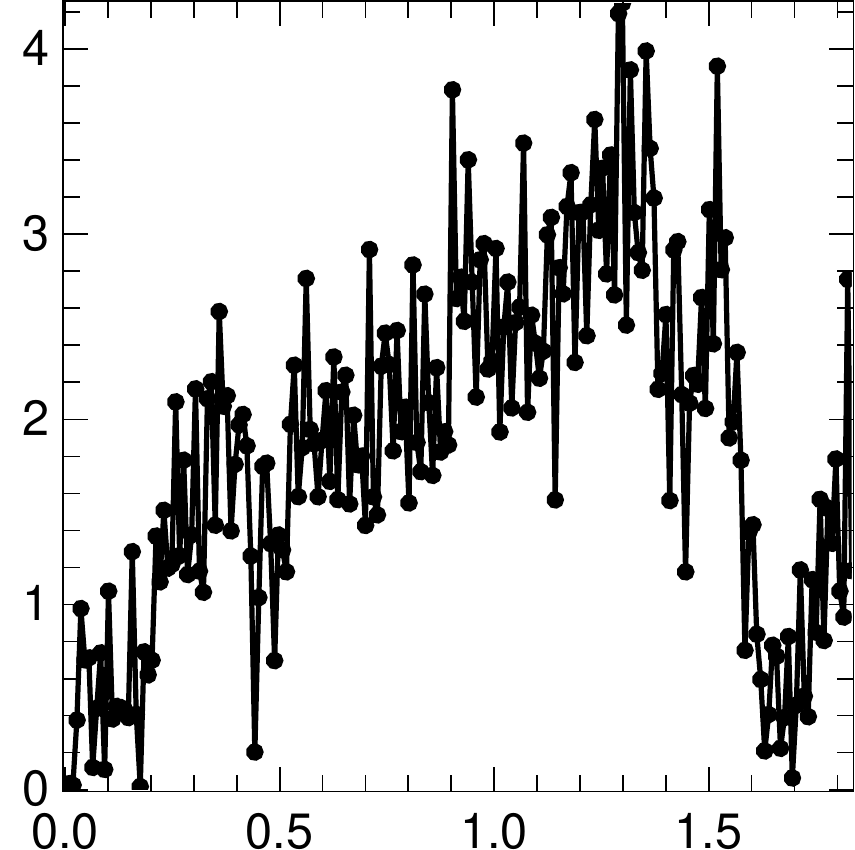}    $L_{\rm {x_0}}$ (m)       
		\end{subfigure}
	\end{tabular} 
	\caption{\label{Chap7:fig4} The absolute error in the lateral pupil position  measurement $|L_{\rm x}-L_{\rm {x_0}}|$  as a function of input lateral pupil shift $L_{x_0}$. The measurements are in the UT scale and where obtained in the laboratory.} 
\end{figure}

The lateral pupil tracking accuracy tests are implemented in two steps by manipulating the PMC actuators.  First, known lateral pupil shifts ($L_{\rm {x_0}}$) are applied to the incoming beams by actuating the PMCs. Second, the input pupil shifts are sensed (say, $L_{\rm x}$) using the pupil tracker function.   Figure~\ref{Chap7:fig4} presents the lateral pupil tracking accuracy as a function of the input pupil shifts. The lateral pupil tracking absolute  accuracy is better than \SI{4}{\milli \meter} at the UT beam magnification (0.05\% of the UT diameter) and the RMS error is \SI{2}{\milli \meter}.

\subsubsection{Longitudinal pupil tracking}

\begin{figure}[h!] 
	\centering	
	\rotatebox{90}{\hspace{-20pt}$|L_{\rm z} - L_{\rm {z_0}}|$ (mm)}
	\begin{tabular}
		{@{}c@{}} 
		\begin{subfigure}{0.28\textwidth}
			\centering
			\includegraphics[width=\textwidth]{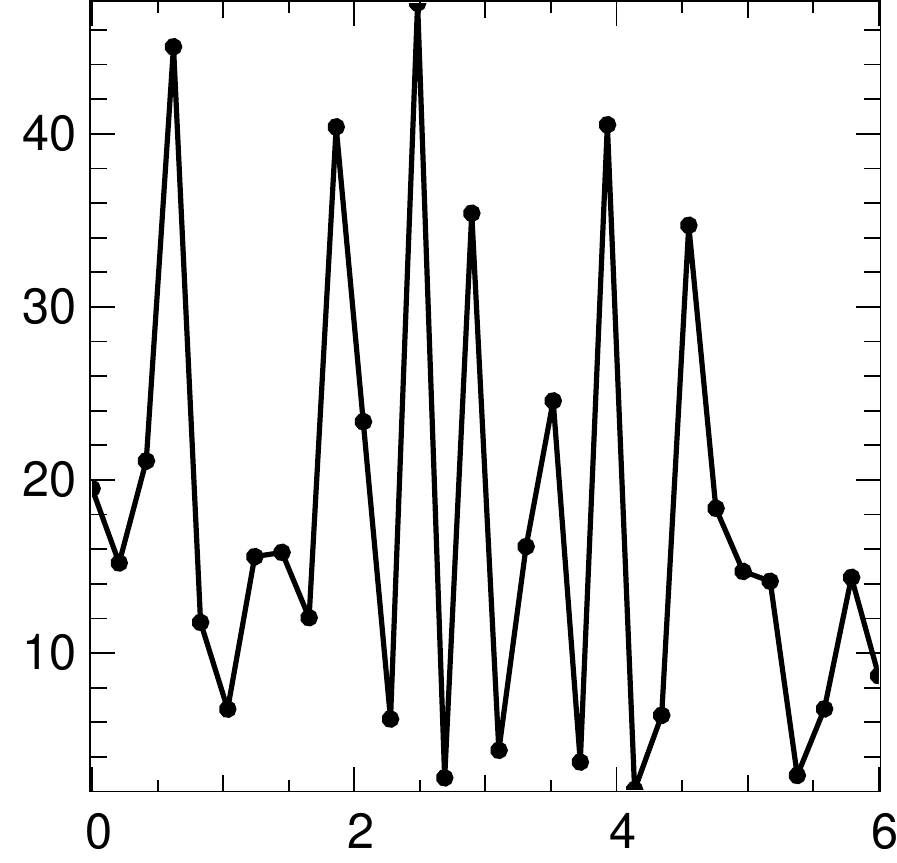}    $L_{\rm {z_0}}$ (meter)       
		\end{subfigure}
		\hspace{11pt }\rotatebox{90}{meter}
		\begin{tabular}
			{@{}c@{}} 
			\begin{subfigure}{0.28\textwidth}
				\centering
				\includegraphics[width=\textwidth]{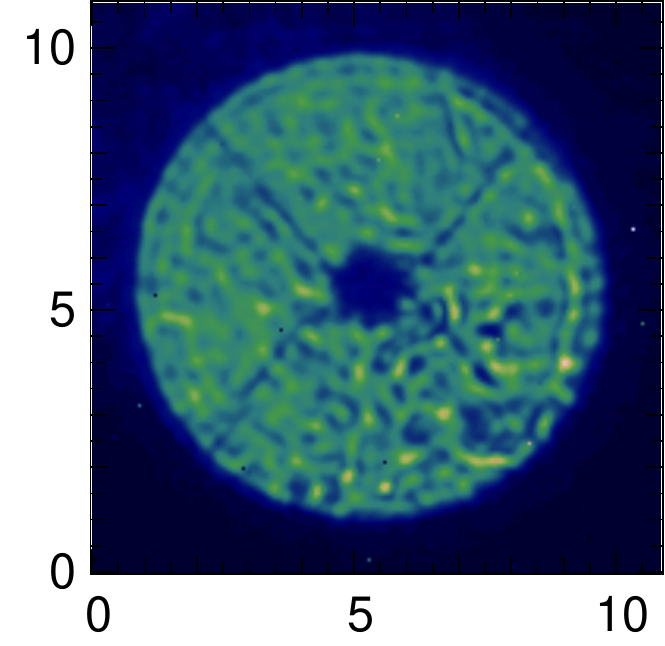}     meter       
			\end{subfigure}
		\end{tabular} \hspace{5pt} \rotatebox{90}{meter}
		\begin{tabular}
			{@{}c@{}} 
			\begin{subfigure}{0.28\textwidth}
				\centering
				\includegraphics[width=\textwidth]{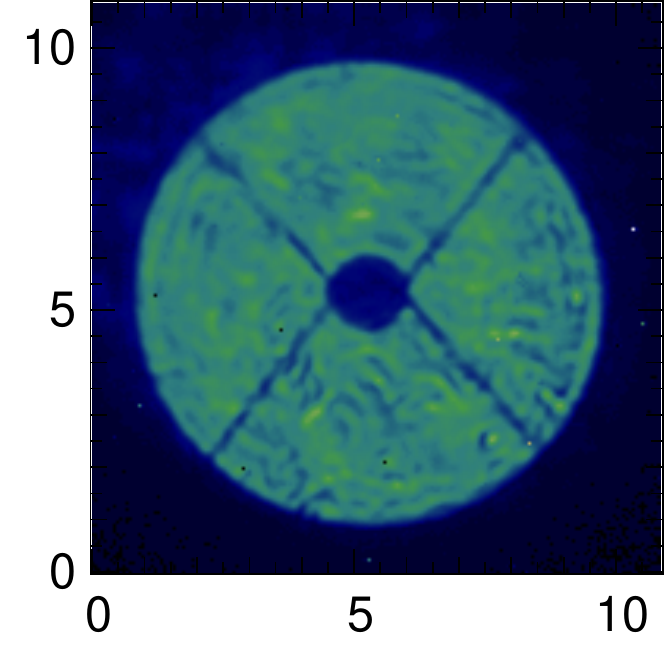}      meter       
			\end{subfigure}
		\end{tabular} 
	\end{tabular}

	\caption{Left: On-sky absolute longitudinal pupil position error  measurement ($|L_{\rm z}-L_{\rm {z_0}}|$) at 80\,mm as a function of input longitudinal pupil position $L_{z_0}$. Center, right: On-sky pupil images obtained before and after closing the pupil guiding closed loop. }
	\label{fig10}
\end{figure}

Longitudinal pupil tracking precision is characterized by manipulating the VCM positions. In this experiment, known longitudinal pupil shifts are applied to the beam by moving the VCM and those are measured back by the pupil tracker. Figure~\ref{fig10} (left panel) presents the longitudinal pupil characterization results and it can be seen that the accuracy is better than 40\,mm for the 80\,mm beam.  The middle and right panel present,  respectively, the UT pupil before and after closing the pupil guiding loop.

\subsection{On-sky beam guiding performance}
The acquisition camera beam guiding has been tested for the ATs and for the UTs. The guiding difference between the ATs and the UTs is that the ATs are not corrected by CIAO. For this reason, the field imager is used to measure slow (tunnel) atmospheric tip-tilts. Whereas for the UTs, the field imager is used to measure the residual tip-tilts of the CIAO. The ATs telescope focus correction is implemented by tracking the focus using the aberration sensor and correcting it with their M2 mirrors.

As stated previously, the GRAVITY field and pupil offset corrections are achieved using two types of actuators: a) the VLTI star separator actuators~\cite{Nijenhuis2008} and the delay line VCM; b) GRAVITY internal field (TTP) and pupil (PMC) actuators. The VLTI star separator has a field selector mirror and a pupil position mirror (M14, laterally and longitudinally movable). Small field and pupil corrections occurring during interferometric observations are corrected in closed loop using GRAVITY internal actuators for the speed and accuracy.  Large field and pupil offsets (which usually take place during the initial alignment of GRAVITY with the VLTI) are corrected using star separator actuators. While operating with GRAVITY internal actuators in the closed loop, when the offsets are larger than GRAVITY internal actuators range, they are offloaded to the VLTI actuators.

During the installation, verification and commissioning phases, the guiding loops of the field, the pupil were extensively tested. Characterization of the field guiding is realized by observing several dual-field stars and checking the stability of injection of star's light into single mode fibers. Characterization of the pupil guiding is implemented by applying known pupil shifts with pupil correcting actuators and checking their stability over an hour. The pupil tracker experiences high backgrounds from bright astrophysical targets due to the closeness of: a) the operating wavelengths of the pupil tracker (operating at 1.2\,$\mu$m) and the field tracker (operates at $H$-band -- 1.65\,$\mu$m) and; b) no adequate dichroic filter for the pupil tracker. The operation of the pupil guiding for an interferometric observation depends on two factors: a) the flux of the pupil guiding laser beacons; b) the magnitude of the astrophysical target. To remove the astrophysical target background, a new software mode, BLINK, is implemented. The basic idea of this  mode is to remove the background by switching OFF and ON the pupil beacons.  During the turn OFF period, the background  is stored and the is subtracted from the subsequent frame. The performance of the pupil guiding is improved by 2 magnitudes when using this BLINK mode. Currently it can operate up to magnitude 1 for the ATs. 

The non-common path errors between the CIAO and  GRAVITY are calibrated by inputting known wavefronts to the GRAVITY beams that are generated by manipulating the MACAO deformable mirror~\cite{Arsenault2003} and measuring the input wavefront aberrations using the aberration sensor. The offsets are then taken into account by CIAO to compensate the non-common path aberrations.

The following summarizes the current on-sky performance of the acquisition camera:

\begin{itemize}
	\item \textbf{Pupil guiding residuals:} The standard deviation of the lateral and the longitudinal pupil guiding residuals are remain smaller than $\pm~0.2$ pixels RMS on the detector (i.e., 12\,mm at the UTs) and the 50\,mm for 80 beam size respectively in the presence of faint astrophysical target.  
	\item \textbf{Field guiding residuals:}  The standard deviation of the field guiding residuals are smaller than $\pm~0.36$ pixels (6.4\,mas). When the tests of the field guiding for the ATs (not equipped with the CIAO) took place, the laser based tunnel seeing tip-tilt tracking system was not  installed. That is why the residuals are large.
	\item \textbf{Focus guiding} The standard deviation of the focus guiding residuals for the ATs are within $\pm~\lambda/8$ RMS.
\end{itemize}

\section{Discussion and conclusions}
The GRAVITY acquisition camera was installed at the Very Large Telescope Interferometer during the last months of 2015. Since then it has been successfully working in the closed loop to stabilize the slow (tunnel) atmospheric turbulence field motions and the pupil motions.  

The camera is characterized using laboratory generated telescope beams at MPE and on-sky at the Paranal Observatory.   The characterization results revealed that it is able to analyze the telescope beams with the  accuracy required by $10\,\mu$as astrometry, namely: a) field tracking with 2\,mas RMS; b) lateral pupil tracking with 4\,mm RMS (at the UT scale); c) longitudinal pupil tracking with 50\,mm RMS (at 80 mm beam scale); and d)  quasi-static higher order wavefront aberration measurements with 80\,nm RMS.

The acquisition camera measured beam parameters are used in the stabilization of the field, the pupil and the focus correction. With the stabilization of the field, the star's light injection into single mode fibers is improved. The quasi-static higher order aberrations measurements are used for the non-common path errors corrections between the CIAO and GRAVITY. The pupil imager is used for visual monitoring. By stabilizing the telescope beams, the astrometric error induced by the field (accuracy of 2\,mas RMS, cf. Section \ref{FI}) and the pupil errors (accuracy of 12\,mm RMS) is minimized to $0.34\,\mu$as).

\section*{Acknowledgments} 
Anugu acknowledges FCT-Portugal grant SFRH/BD/52066/2012.   The research leading to these results was partially supported by FCT-Portugal grant PTDC/CTE-AST/116561/2010 and the European Community's Seventh Framework Programme under Grant Agreement 312430.


\bibliography{References.bib}   
\bibliographystyle{spiebib}   
\end{document}